\documentclass[fleqn,showpacs,showkeys,twocolumn,nofootinbib]{revtex4}
\usepackage{graphicx}
\usepackage{amssymb}
\usepackage{amsmath}
\usepackage{bm}

\newcommand{\ds}{\displaystyle}
\newcommand{\dsf}{\ds\frac}

\begin{document}

\setcounter{page}{0}
\title[]{Neutron star structure in an in-medium modified chiral soliton model}
\author{U.~T.~Yakhshiev}
\affiliation{Department of Physics, Inha University, Incheon
402-751, Republic of Korea }


\begin{abstract}
We study the internal structure of a static and spherically symmetric 
neutron star
in the framework of an in-medium modified chiral soliton model. 
The Equations of State describing an infinite and asymmetric nuclear matter
are obtained introducing the density dependent functions into the low 
energy free space Lagrangian of the model starting from the phenomenology 
of pionic atoms. 
The parametrizations of density dependent functions are related to the 
properties of isospin asymmetric nuclear systems at saturation density 
of symmetric nuclear matter $\rho_0\simeq 0.16$~fm$^{-3}$. 
Our results, corresponding to the compressibility of symmetric nuclear matter 
in the range $250\,\mbox{MeV}\le K_0\le 270\,\mbox{MeV}$ and the  slop parameter
value of symmetry energy in the range
$30\,\mbox{MeV}\le L_S\le 50\,\mbox{MeV}$,  are consistent  with the results from 
other approaches and with the experimental indications.
Using the modified Equations of State, 
near the saturation density of symmetric nuclear matter $\rho_0$,
 the extrapolations 
to the high density and highly isospin asymmetric regions have been performed. 
The calculations showed that the properties of $\sim 1.4M_\odot$ and 
$\sim 2M_\odot$  neutron stars 
can be well reproduced in the framework of present approach.
\end{abstract}




\maketitle

\section{Introduction}

Analysis of the neutron star structure is an interesting 
topic of the modern astrophysics. In particular, in the 
nuclear astrophysics the structure studies of neutron stars 
are related to the Equations of State (EOS) of nuclear matter 
which describes the pressure density versus 
energy density relation for a broad range of density values
(for example, see recent review~\cite{Lattimer:2012nd} and references therein).
The peculiarities of EOS are well know at the densities below 
the saturation density of symmetric nuclear matter $\rho_0$ while at
the high density regions they are still remaining not clear. 
The high density behavior of EOS are 
poorly understood because of the difficulty of direct experimental accessibility
in laboratories and because of 
the absence of {\em ab initio} theoretical calculations. 
Therefore, from the experimental point of view, the neutron star studies 
may serve as a laboratory for understanding the behavior of EOS at high densities.
From the theoretical point of view, instead of {\em ab initio} calculations one can 
start from the phenomenological framework taking into account the well known 
properties of EOS at the low density region and extrapolate to the high density
regions trying to describe the properties of matter under the extreme 
(high density and high temperature) conditions.

In this context and if one able to formulate further in-medium modifications,
a chiral soliton model of Skyrme,
describing the single nucleon properties in free space~\cite{Skyrme:1961vq,Adkins:1983ya}, 
or its variations including the explicit 
vector mesonic degrees of freedom~\cite{Meissner:1987ge} 
may serve as a starting point for the theoretical framework. The in-medium modifications
may be expressed allowing the density dependencies of the constants 
entering into the initial free space Lagrangian. It is necessary to note, 
that in principle one should be able also to reproduce 
the medium dependencies in the effective Lagrangian 
starting from the first principles but it is not known yet the form of general 
low energy Lagrangian and the peculiarities of its ingredients. 
For this reason, in the present work we use an in-medium modified 
chiral soliton model~\cite{Yakhshiev:2014hza,Yakhshiev:2013eya} which 
may be considered as a truncated 
version of the general low energy Lagrangian.

In Refs.~\cite{Yakhshiev:2014hza,Yakhshiev:2013eya} the medium modifications were 
achieved putting a bit more phenomenology into the initial model, i.e. putting 
the density dependent functions into the free space Skyrme Lagrangian
according to the pionic-atoms data at low energies~\cite{Ericsonbook}
 and properties of asymmetric nuclear 
matter at saturation density $\rho_0$. Although the in-medium modified 
Skyrme Lagrangians is assumed to be  very truncated version of the possible 
general Lagrangian, it must be applicable  
to the studies of nuclear many-body problems in the spirit of chiral effective 
Lagrangians. The pay for the truncation may be the possible deviations from 
the experimental observables in the sense of quantitative description. 
Nevertheless, the model has obvious virtues: i) it has the simplest 
Lagrangian among the same class Lagrangians, and ii) it seams have all 
necessary ingredients for the qualitative description of the  
strong interaction physics. These ideas were the basic ruling 
philosophy behind of 
the approach developed in Refs.~\cite{Yakhshiev:2014hza,Yakhshiev:2013eya}
and we continue our model studies in the present work. 

The model is phenomenological one and must pass as much as possible tests
on its applicability to strong interacting systems
comparing with other approaches and the 
experimental indications.
The previous nuclear matter studies~\cite{Yakhshiev:2013eya} showed that 
the in-medium modified Skyrme term is responsible for preventing the collapse 
of nuclear matter to the singularity at high densities in analogy 
to the free space case where the Skyrme term is responsibly for the 
stabilization of finite size solitons. The modifications showed that at 
some values of model parameters the properties of infinite and isospin 
asymmetric nuclear matter can be reproduced well near the saturation 
point of symmetric nuclear matter $\rho_0$.

It will be interesting to test the applicability of model to the 
strong interacting systems under the extreme conditions 
extrapolating the modified Equations of State to the high density regions.
Therefore, in the present work we make further check of the 
basic philosophy considering an
application of the model for the studies of neutron stars structure.

\section{Formalism}

\subsection{The Lagrangian}
\label{subsec:modelLag}

Our starting point is the in-medium modified Skyrme-model
Lagrangian described in Ref.~\cite{Yakhshiev:2014hza}
\begin{eqnarray}
{\cal L}^*&=&{\cal L}_2^*+{\cal L}_4^*+{\cal L}_{m}^*
+{\cal L}^*_{e}\,,
\label{lmed}\\
{\cal L}_2^*\!&=&\!\dsf{F_{\pi}^{2}}{16}\,\alpha_\tau
\mbox{Tr}\left(\partial_0U\partial_0U^\dagger\right)
-\dsf{F_{\pi}^{2}}{16}\,\alpha_s\mbox{Tr}\left(\partial_i
U\partial_iU^\dagger\right),\quad\cr
{\cal L}_4^*&=&
-\dsf{1}{16e^2\zeta_\tau}\,\mbox{Tr}\,\left[U^\dagger\partial_0 U,
U^\dagger \partial_i U\right]^2\nonumber\\
&&+\dsf{1}{32e^2\zeta_s}\,\mbox{Tr}\,\left[U^\dagger\partial_i U,
U^\dagger \partial_j U\right]^2\,,\cr
{\cal L}_{m}^*&=&-\dsf{F_{\pi}^{2} m_{\pi}^{2}}{16}\,\alpha_m
\mbox{Tr}\,\left(2-U-U^+\right)\,,\cr
{\cal L}_{e}^*\!&=&\!-\dsf{F_\pi^2}{16}\, m_\pi\alpha_e
\varepsilon_{ab3}
\mbox{Tr}\left(\tau_a U\right)\mbox{Tr}\left(\tau_b\partial_0
U^\dagger\right),\nonumber
\end{eqnarray}
where Einstein's summation convention is always assumed (if not specified otherwise).
The chiral $SU(2)$ matrix $U=\exp(2i \tau_a \pi_a/F_\pi)$
is defined in terms of the
Cartesian isospin-components of the pion field $\pi_a$ ($a=1,2,3$).
The density functionals entering into the Lagrangian
($\alpha_\tau$, $\alpha_s$, $\zeta_\tau$, $\zeta_s$, $\alpha_m$ 
and $\alpha_e$) represent the influence of
surrounding environment to the single soliton properties.

In the single skyrmonic sector, the input parameters of the model have
the following values:
$F_\pi=108.783$\,MeV, $e=4.854$ and $m_\pi=134.976$\,MeV.
They correctly reproduce the
experimental values of the nucleon mass $m_N=938$\,MeV and 
$\Delta$-isobara mass $m_\Delta=1232$\,MeV
in free space, i.e. if
$\alpha_\tau=\alpha_s=\alpha_m=\zeta_\tau=\zeta_s=1$ and $\alpha_e=0$.

This simple Lagrangian describes the properties of nucleons in free space,
the properties of nucleons in nuclear matter as well as the properties of 
isospin asymmetric nuclear matter starting at the same footing.
The formalism for the classical 
solitonic solutions, the quantization method 
and the explanations of applications for symmetric and asymmetric matter properties 
can be found in Refs.~\cite{Yakhshiev:2014hza,Yakhshiev:2013eya}
and references therein. But for seflconsistency,
in the next subsection~\ref{subsec:nuclmat} we briefly 
outline how an infinite size nuclear systems can be described
in the framework of the present approach and represent    
some final formulas.

\subsection{Nuclear matter}
\label{subsec:nuclmat}

In the thermodynamic limit at zero temperature, for a system of 
an infinite number of baryons uniformly distributed in infinite volume
but keeping the density per unit volume constant, the binding
energy per nucleon can be represented as
\begin{equation}
\varepsilon(\lambda,\delta)=\varepsilon_V(\lambda)
+\varepsilon_A(\lambda,\delta)\,,
\label{be}
\end{equation}
where $\varepsilon_V$ and $\varepsilon_A$ are known as volume and
asymmetry energies, respectively.
For the convenience, here we introduced the 
isoscalar $\lambda=\rho/\rho_0$ and isovector
$\delta=\delta\rho/\rho$ density parameters 
in terms of the isoscalar $\rho=\rho_{\rm neutron}+\rho_{\rm proton}$  
and isovector $\delta\rho=\rho_{\rm neutron}
-\rho_{\rm proton}$ nuclear densities, 
and the normal nuclear matter density $\rho_0$.

In the framework of present approach, 
using the spherically symmetric approximation for a single 
soliton properties via the hedgehog ansatz
$U=\exp\{i\tau_ir_iF(r)/r\}$ and considering an isospin asymmetric
and infinite size nuclear environment,
one can get the symmetric and asymmetric parts 
of the binding energy per nucleon~\cite{Yakhshiev:2014hza}
\begin{eqnarray}
\varepsilon_V(\lambda)&=&
f_1m(f_2)+\frac{3f_3}{8\Lambda}-m_N\,,\\
\varepsilon_A(\lambda,\delta)
&=&\frac{\Lambda_2}{2\Lambda}
\left(1+\frac{\Lambda_2m_\pi f_4}{f_3}\right)m_\pi f_4\delta^2.\,\,\,\quad
\end{eqnarray}
Here the functionals $m$, $\Lambda$ and $\Lambda_2$  are defined as 
\begin{eqnarray}
m(f_2)&=&\frac{\pi F_\pi}{e}
\int\limits_0^\infty dx\,x^2
\Big\{f_2\frac{2m_\pi^2}{e^2F_\pi^2}(1-\cos F)\cr
&+&\frac{F_x^2}2+\dsf{2\sin^2F}{x^2}
\left[\frac12+2F_x^2+\frac{\sin^2F}{x^2}\right]\Big\},\qquad\\
\Lambda_2&=&\frac{2\pi}{3e^3F_\pi}
\int\limits_0^\infty x^2 \sin^2F\, dx\,,
\label{LF2}\\
\Lambda_4&=&\frac{8\pi}{3e^3F_\pi}
\int\limits_0^\infty \left(F_x^2+\frac{\sin^2F}{x^2}\right) x^2
\sin^2F\, dx\,.
\label{LF4}
\end{eqnarray}
In the functionals above, using an infinite nuclear matter approximation, 
the initial medium functionals in Eq.~(\ref{lmed}) were rearranged by defining 
the new medium functions $f_{1,2,3,4}$.  The rearrangements
are made in the following way  
\begin{eqnarray}
1+C_1\rho&=f_1\equiv&\sqrt{\frac{\alpha_s}{\zeta_s}}\,,\\
1+C_2\rho&=f_2\equiv&\frac{\alpha_m}{(\alpha_s)^2\zeta_s}\,,\\
1+C_3\rho&=f_3\equiv&\frac{(\alpha_s\zeta_s)^{3/2}}{\alpha_\tau}\,,\\
\frac{C_4\lambda}{1+C_5\lambda}\delta&=f_4\equiv&\frac{\alpha_e}{\alpha_\tau}\delta^{-1}
\end{eqnarray}
and the basic principle behind the linear density dependent 
parametrizations of functions $f_i$ was the simplicity of form.

In the last four equations $C_i$ ($i=\overline{1,5}$) 
are the model parameters and, therefore, the present 
approach is a simple 5-parametric solitonic model of nuclear 
matter.\footnote{We would like to note, 
although at present work we discuss the properties 
of an asymmetric and infinite nuclear matter the model can be applied 
for the studies of finite nuclei properties
after fitting the density parameters $C_i$.}
As soon as we define 5-parameters at some specific density (e.g. the normal 
nuclear matter density, $\rho_0$) the Equations of State for the symmetric 
and asymmetric nuclear matter are defined for any values of nuclear matter density 
if one assumes that the extrapolations by means of the linear on 
density functions $f_i$ 
are valid.\footnote{Other details can be found in 
Refs.~\cite{Yakhshiev:2014hza,Yakhshiev:2013eya}.}
In the next subsection~\ref{subsec:neutstar} we concentrate our attention 
on the properties of baryonic systems at high densities discussing 
the neutron stars.

\subsection{Neutron star}
\label{subsec:neutstar}

As the first approximation in describing the neutron stars,
one can consider a spherically symmetric and static mass distribution.
Then any part of neutron star mass ${\cal M}(r)$ inside a 
sphere with radius $r$ is given by the integral
\begin{eqnarray}
{\cal M}(r)&=&4\pi \int_0^r {\rm d}r\,r^2 {\cal E}(r)\,.
\label{Mr}
\end{eqnarray}
Here ${\cal E}(r)$ is the mass-energy density distribution of the 
neutron star in the radial direction. 
Consequently, the total 
gravitational mass $M$ of the neutron star with radius $R$ is 
defined by the condition
\begin{equation}
M={\cal M}(r\ge R)\,.
\end{equation}
Further, in the spherically symmetric approximation, the pressure 
density change in radial direction inside the neutron star is given by 
 the Tolman-Oppenheimer-Volkoff (TOV) 
 equation~\cite{Tolman:1939jz,Oppenheimer:1939ne}
\begin{eqnarray}
-\frac{dP(r)}{dr}&=&\frac{G{\cal E}(r){\cal M}(r)}{r^2}\left(1-
\frac{2G{\cal M}(r)}r\right)^{-1}\cr
&\times&\left(1+\frac{P(r)}{{\cal E}(r)}\right)\left(1+
\frac{4\pi r^3P(r)}{{\cal M}(r)}\right)\,.
\label{Pr}
\end{eqnarray}
Here $G=6.707\times10^{-39}\hbar c\left(\frac{\rm GeV}{c^2}\right)^{-2}$
is gravitational constant and 
$P(r)$ is pressure density in radial direction.
The boundary conditions for the functions entering into the equation are 
\begin{equation}
{\cal M}(0)=0\qquad \mbox{and}\qquad {\cal E}(0)={\cal E}_{\rm cent}\,.
\end{equation}
After solving TOV equation one can find 
the radius $R$ of a star with central energy density ${\cal E}_{\rm cent}$.
It is defined by the pressure zero condition 
at the surface of the neutron star, $P(r=R)=0$. 

To obtain the numerical solution 
for the profile of a star, one solves Eqs.~(\ref{Mr}) and~(\ref{Pr})
using the Equation of State 
\begin{eqnarray}
P&=&P({\cal E})\,.
\end{eqnarray}

To find $P({\cal E})$ dependence in present approach, we note that 
the pressure and energy dependencies on the  
density parameter $\lambda$ for the neutron matter ($\delta=1$) 
are given by equations
\begin{eqnarray}
\label{Pden}
P(\lambda)& =& \rho_0\lambda^2\frac{\partial \varepsilon(\lambda,1)}{\partial\lambda}\,,\\
{\cal E}(\lambda) &=& [\varepsilon(\lambda,1)+m_N]\lambda\rho_0.
\label{Eden}
\end{eqnarray}
where $\varepsilon(\lambda,1)$ is binding energy per nucleon in 
neutron matter. The system of parametric equations,
Eq.~(\ref{Pden}) and Eq.~(\ref{Eden}),
 gives the desired relation $P({\cal E})$
between pressure and energy densities in a neutron star. 

\section{Results and discussions}

In general, the thermodynamically limiting binding energy per 
nucleon given in Eq.~(\ref{be}) must be valid at any densities 
and at any given isospin parameter $\delta$.
For example, it is valid also at the extreme conditions which are 
formed in interior regions of neutron stars. 
Those extreme conditions described by very high density 
($\rho\simeq$ several times of $\rho_0$) and
highly isospin asymmetric ($\delta\simeq 1$) form of nuclear matter. 

From other side, 
it is not clear the direct relation of the liquid-drop 
formula of Bethe and 
Weizs\"{a}cker~\cite{Weizscker:1935zz,Bethe:1936zz}
\begin{equation}
B(Z,N)=a_VA-a_SA^{2/3}-a_C\frac{Z^2}{A^{1/3}}-a_A\frac{(N-Z)^2}{A}+...\,,
\label{BW}
\end{equation}
describing the binding energy of nucleus to the binding energy 
of neutron star. The reason is not only due to the 
relativistic metric factors 
coming from the Einstein's equations in the calculations of 
binding energies of neutron stars.
The reason here is rather obvious, the semiempirical 
liquid-drop formula describes the binding energy 
per nucleon near the normal nuclear matter densities 
(which correspond to the
density profiles of the  existing heavy nuclei)
and its validity at high densities is not clear.

Nevertheless, 
around the normal nuclear matter density and for the small values of 
asymmetry parameter $\delta$, 
the thermodynamically limiting binding energy per 
nucleon given in Eq.~(\ref{be}) is well related 
to the Bethe and  Weizs\"{a}cker's formula. 
Because, from one side, 
in the limit of small $\delta$ 
the binding energy formula in Eq.~(\ref{be}) can 
be approximated as 
\begin{equation}
\varepsilon(\lambda,\delta)=\varepsilon_V(\lambda)
+\varepsilon_S(\lambda)\delta^2+{\cal O}(\delta^4)\,,
\label{symen}
\end{equation}
where $\varepsilon_S(\lambda)$ is called the symmetry energy.
From other side, 
if one ignores the Coulomb and surface effects, 
the binding energy per nucleon defined from the liquid-drop formula 
will take the form
\begin{equation}
A^{-1}B(Z,N)\approx a_V-a_A\delta^2+...\,.
\end{equation} 
From the comparisons of the last two equations, 
it is seen that the density dependencies of volume and 
symmetry energies can be established well 
at the densities around $\rho_0$ from the 
stability conditions and the density variations of 
resonating heavy nuclei near the ground state.

The symmetry energy ${\varepsilon_S}$
describes the energy increase in the system if 
the number of protons and neutrons becomes not equal relatively 
to the case when the neutron and proton numbers are same. 
Consequently, the symmetry energy is important factor
in describing the properties 
of neutron-reach stable nuclei existing in nature as well as 
the properties of exotic nuclei formed under extreme conditions where the 
neutron-to-proton number $N/Z$ is much smaller or much larger than 
one~\cite{Baran:2004ih}.
Although the definition of symmetry energy
in Eq.~(\ref{symen}) is model independent
its density dependence is clearly model dependent. 
Therefore,  
during the fitting to density region near the normal nuclear matter density
$\rho_0$ the different models are modified taking into account 
the properties of symmetry energy coming from the 
phenomenological observations. 
But the extrapolations of EOS to the high density 
($\rho>>\rho_0$) and highly asymmetric ($\delta\simeq 1$) regions 
remain not clear.
In particular, this is due to the reason that 
in neutron stars the higher 
order terms in $\delta$ in Eq.~(\ref{symen}) may be also important. 
Usually, it is assumed that ${\cal O}(\delta^4)$ 
terms are small and the symmetry energy is mostly 
responsible in describing the properties of neutron stars
when the neutron-to-proton number $N/Z$ becomes 
infinite.\footnote{Although the neutron star is initially formed from 
the ordinary matter which has the finite neutron-to-proton number
with well separated electrons, 
due to the gravitational collapse and due to further 
formed electron degenerate states, 
and due to the following nucleon degenerate states at even higher 
densities the frequent collisions lead to the intensive nuclear 
reactions. As a result "an effective" neutron-to-proton number $N/Z$ becomes 
infinite.} In the present approach, the higher order terms 
${\cal O}(\delta^4)$ come from the 
explicit isospin breaking symmetry in the mesonic sector and found to be 
negligible.\footnote{See the discussions in subsection 
V.C of Ref.~\cite{Yakhshiev:2013eya}.} 
Consequently, we ignore them in the present work.

\subsection{Nuclear matter and Symmetry energy}

\begin{figure}
\vspace{0.2cm}
\hspace{.cm}
\centerline{\includegraphics[width=8cm]{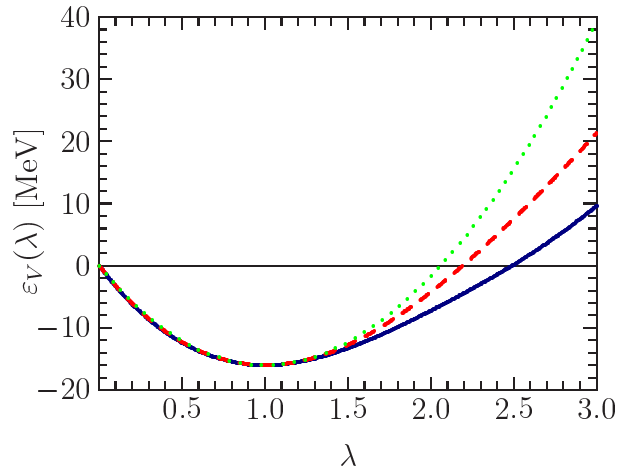}}
\caption{(Color online) The volume energy
as a function of normalized density $\lambda=\rho/\rho_0$.
Solid, dashed and dotted curves correspond to the values of
compressibility $K_0=230$\,MeV,
$K_0=250$\,MeV and $K_0=270$\,MeV, respectively.
Other parameters are defined in Table~\ref{table1}.}
\label{fig:SM}
\end{figure}

Let us first discuss the properties of the symmetric nuclear matter.
The value of the first experimental parameter in the liquid-drop 
formula Eq.~(\ref{BW}) is well known, $a_V\approx 16$\,MeV. 
It is also well established that the saturation density of 
symmetric nuclear matter,
where the pressure becomes zero $P_{\rm sym}=0$, is around 0.16\,fm$^{-3}$.
Further, the compressibility of symmetric nuclear matter within the various
approaches is found to be $K_0\sim 290\pm 70$\,MeV at 
the saturation density $\rho_0$~\cite
{Sharma:1988zza,Shlomo:1993zz,Ma:1997zzb,Vretenar:2003qm,TerHaar:1986ii,Brockmann:1990cn}.
Therefore, these three quantities may be used to fix some of the 
model parameters (e.g. $C_1$, $C_2$ and $C_3$)
if one expands the volume energy into the Taylor series
\begin{equation}
\varepsilon_V(\lambda)=\varepsilon_V(1)+\frac{K_0}{18}(\lambda-1)^2
+\dots
\end{equation}
around the saturation density, $\lambda=1$. 
After fitting the values of parameters, the EOS of symmetric matter
is defined for any values of density.\footnote{
In this case, the minimization scheme is related to the single 
in-medium soliton, i.e. the soliton mass is minimized at the given 
values of density functions $f_i$.} 
The density dependencies of the binding energy per nucleon 
in symmetric matter are shown in Fig.~\ref{fig:SM}
for the three possible values of the compressibility $K_0$.
The corresponding parameters and 
the volume energy coefficients at saturation density $\rho_0$
are given in Table~\ref{table1}.

It is seen, that our results become consistent with the results from other approaches
and the phenomenological indications at some values of 
density parameters $C_i$.
For example, the density dependence of volume energy from the Set II is 
close to the APR (Akmal-Pandharipande-Ravenhall) predictions~(see 
the model A18$+\delta v+$UIX$^*$ in Ref.~\cite{Akmal:1998cf})
made on the basis of Argonne $v_{18}$ two-nucleon interactions~\cite{Wiringa:1994wb}.
Similarly, the results from the Set III is very close to the results from the
model A18$+$UIX~\cite{Akmal:1998cf}.

\begin{table}[hbt]
\caption{Three sets of parameters which reproduce
the symmetric matter properties. The 
parameters $C_1$, $C_2$ and $C_3$ are chosen in
such a way that at saturation point $P_{\rm sym}(\rho_0)=0$
the value of volume energy per nucleon equals to its
experimental value, $\varepsilon_V(\rho_0)=
\varepsilon_V^{\rm exp}\approx-a_V$, and the compressibility of nuclear
matter $K_0$ is reproduced in a given experimental range.}
\begin{ruledtabular}
\begin{tabular}{ccccccc}
Set
&$C_1$
& $C_2$
& $C_3$
& $\varepsilon_V(\rho_0)$
& $K_0$
& $Q$ \\
&&&
& [MeV]
& [MeV]
& [MeV]\\
\colrule
I   &  $-0.285$  &0.803 &  1.753  & $-16$ & 230 & $-545$ \\
II  &  $-0.273$  &0.643 &  1.858  & $-16$ & 250 & $-279$ \\
III &  $-0.333$  &0.281 &  3.090  & $-16$ & 270 & $-133$ \\
\end{tabular}
\end{ruledtabular}
\label{table1}
\end{table}

The quantity in the last column, referred as the skewness 
parameter, is proportional to the third derivative
of the volume term at saturation density $\rho_0$ and defined as
\begin{equation}
Q=27\lambda^3\,\left.\frac{\partial^3\varepsilon_V(\lambda)}
{\partial\lambda^3}\right|_{\lambda=1}\,.
\end{equation}
Its values presented in Table~\ref{table1} are outcome results 
from the present approach. 
There is a nonlinear correlation between $Q$ and compressibility $K_0$,
$Q$ increases if the value of $K_0$ increases.
Our predictions for $Q$ is qualitatively similar to the results from the Hartree-Fock
approach based on Skyrme interactions~\cite{Chabanat:1997qh}, and
to the result from the MDI
(isospin and momentum-dependent interaction) model~\cite{Das:2002fr}.
Another example, the phenomenological momentum-independent model (MID)
also predicts the similar results~\cite{Chen:2009qc}. For comparison,
in the present model one has $Q/K_0\approx-1.71$ at $K_0=240$\,MeV while MID model
gives the result $Q/K_0\approx-1.6$ at that value of $K_0$.

Now let us discuss the properties of the asymmetric matter. 
As we said above, 
in the approximation that the higher order terms in $\delta$ 
are small (see Eq.~(\ref{symen})), the asymmetric matter properties are 
completely determined by the density dependence of the symmetry energy.
The properties of the symmetry energy can be studied again
by expansion into the Taylor series  
\begin{equation}
\varepsilon_S(\lambda)=\varepsilon_S(1)+
\frac{L_S}3(\lambda-1)+\frac{K_S}{18}(\lambda-1)^2
+\dots
\label{eq:SE}
\end{equation}
around the saturation density, $\lambda=1$.
While the first coefficient $\varepsilon_S(1)$ in Eq.~(\ref{eq:SE})
is known to be more or less well defined phenomenologically,  
$\varepsilon_S(1)\sim 29$ to 34~MeV, the values of slop parameter 
$L_S$ and compressibility of asymmetric matter $K_S$ remain unclear. 
The reason is that EOS of asymmetric matter is highly sensitive to 
those parameters and the different models give the different predictions.
For example,
in relativistic mean field approaches (see Ref.~\cite{Agrawal:2012rx},
for the recent optimized versions) there are mainly two classes: i)
small $\varepsilon_S(\rho_0)\sim 30$\,MeV and small $L_S\sim50$\,MeV (BSP, IUFSU$^*$, IUFSU)
and ii) large $\varepsilon_S(\rho_0)\sim37$\,MeV and large $L_S\sim110$\,MeV (G1, G2, TM1$^*$, NL3).
The recent analyses of the data from heavy-ion collisions~\cite{Tsang:2011ju} and
neutron-skin experiments~\cite{Tsang:2012se} give the prediction
$L_S\sim 70\pm 20$~MeV. 

During the modifications of EOS in the present work we choose 
the value of slop parameter $L_S$ in the interval 30$-$50\,MeV
which is in agreement with various empirical 
constraints~\cite{Tsang:2011ju,Tsang:2012se,Horowitz:2014bja}. 

\begin{figure}[htb]
\vspace{0.2cm}
\hspace{.cm}
\centerline{\includegraphics[width=8cm]{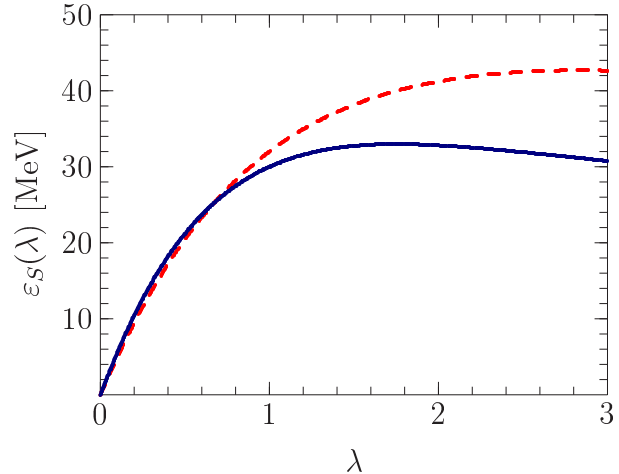}}
\caption{(Color online) The symmetry energy
as a function of normalized density $\lambda=\rho/\rho_0$.
The solid and the dashed curves correspond to the sets 
II-a and III-f defined in Table~\ref{table2}.}
\label{fig:SE}
\end{figure}

The symmetry energies, calculated using two different sets of parameters,
are shown in Fig.~\ref{fig:SE}
and the parameters are determined in Table~\ref{table2}.
\begin{table*}[hbt]
\caption{The different sets of parameters which reproduce
the asymmetric matter properties. The 
parameters $C_4$ and $C_5$ are chosen in
such a way that, at saturation density $\rho_0$,
the value of symmetry energy $\varepsilon_S$ and the coefficient of 
its first derivative $L_S$ are
reproduced in the commonly adopted range. Other parameters
are defined in Table~\ref{table1} (see the parameters values 
corresponding to the same $K_0$ values given in this table).}
\begin{ruledtabular}
\begin{tabular}{ccccccccccc}
Set&$K_0$
&$C_4$
& $C_5$
& $\varepsilon_S(\rho_0)$
& $L_S$
& $K_S$
& $K_\tau$
& $K_{0,2}$
&$\varepsilon_S$(0.1\,fm$^{-3}$)
&$\varepsilon_S$(0.11\,fm$^{-3}$)\\
&[MeV]
&
&
&[MeV]
&[MeV]
&[MeV]
&[MeV]
&[MeV]
&[MeV]
&[MeV]\\
\colrule
II-a &  250 & 2.338  & 0.878 & 30 & 30 & $-209$ & $-299$ & $-265$ & 24.26 &25.54\\
II-b &  250 & 1.984  & 0.594 & 30 & 40 & $-209$ & $-329$ & $-285$ & 23.11 &24.55\\
II-c &  250 & 1.723  & 0.384 & 30 & 50 & $-197$ & $-347$ & $-291$ & 22.06 &23.64\\
II-d &  250 & 2.559  & 0.946 & 32 & 30 & $-222$ & $-312$ & $-278$ & 26.11 &27.44\\
II-e &  250 & 2.183  & 0.660 & 32 & 40 & $-226$ & $-346$ & $-301$ & 24.93 &26.44\\
II-f &  250 & 1.904  & 0.448 & 32 & 50 & $-217$ & $-367$ & $-311$ & 23.86 &25.50\\
III-a &  270 & 2.670  & 1.498 & 30 & 30 & $-169$ & $-259$ & $-245$ & 24.57 &25.76\\
III-b &  270 & 2.179  & 1.024 & 30 & 40 & $-177$ & $-297$ & $-278$ & 23.33 &24.71\\
III-c &  270 & 1.831  & 0.701 & 30 & 50 & $-172$ & $-322$ & $-298$ & 22.22 &23.75\\
III-d &  270 & 2.980  & 1.622 & 32 & 30 & $-178$ & $-268$ & $-254$ & 26.46 &27.69\\
III-e &  270 & 2.425  & 1.133 & 32 & 40 & $-189$ & $-309$ & $-290$ & 25.20 &26.62\\
III-f &  270 & 2.044  & 0.798 & 32 & 50 & $-188$ & $-338$ & $-313$ & 24.05 &25.64\\
\end{tabular}
\end{ruledtabular}
\label{table2}
\end{table*}
Here we present only two representatives among the many sets 
producing the symmetry energy parameters in commonly adopted range.
Depending on the compressibility $K_0$ value of the symmetric nuclear matter
we classify the results into two models, Model II 
with relatively smaller value of the compressibility 
$K_0=250$~MeV (more soft EOS) and 
Model III with relatively bigger value of 
the compressibility $K_0=270$~MeV (more stiff EOS).

The results show that the symmetry energy is less sensitive to the
different model parameters presented in Table~\ref{table2}.
In the Fig.~\ref{fig:SE} we have shown the density dependencies of 
symmetry energy corresponding to two boundary regions referred as
more soft (II-a) and more stiff (III-f) Equations of State.  
The other sets reproduce the density dependence of symmetry energy 
corresponding to somewhere in between of these two boundary curves.
The relatively more stiff EOS (III-f) in the present approach reproduce 
the density dependence of symmetry energy which is  
close to the APR predictions~\cite{Akmal:1998cf}, to the result from 
the MDI model with the parameter $x=0$~\cite{Das:2002fr}
as well as to the results from MID model
with the parameter $y=-0.73$~\cite{Chen:2009qc}.

The quantities in the last five columns in Table~\ref{table2} are outcome 
from our model calculations. In particular,  
the quantities $K_\tau$ and $K_{0,2}$ are related to the compressibility 
of asymmetric matter and defined as
\begin{equation}
K_\tau=K_S-6L_S\,,\qquad K_{0,2}=K_\tau-\frac{Q}{K_0}\,L_S\,.
\label{SEquant}
\end{equation}
They describe the correlations between the symmetry energy coefficients
and important for estimating the shift in compressibility value 
in asymmetric matter 
\begin{equation}
K(\rho_0,\delta)=K_0 + K_{0,2}\delta^2+{\cal O}(\delta^4)\,.
\end{equation}
The condition for the lowering of saturation density value 
in asymmetric matter leads to the 
constraint $K_\tau<0$. This constraint is fulfilled in the present work.

The calculated values of $K_\tau$ and $K_{0,2}$ 
are consistent with the results from other approaches. For example,
the phenomenological momentum-independent model
predicts the range for the values of $K_{0,2}$:
$-477~\mbox{MeV}\le K_{0,2}\le -241~\mbox{MeV}$~\cite{Chen:2009qc}.
Our predictions, made using the different sets, are also belong to 
that range (see Table~\ref{table2}).

It is interesting also to compare the low density behavior of
the symmetry energy in the present model with other model predictions.
For example, an analysis of the giant dipole resonance
(GDR) of ${}^{208}$Pb using a series of microscopic Hartree-Fock plus 
Random phase approximation calculations
predicts the following values of symmetry energy at subnormal nuclear matter density:
$20\,\mbox{MeV}<\varepsilon_S(\rho=0.1\mbox{\,fm}^{-3})<25.4
\,\mbox{MeV}$~\cite{Trippa:2008gr}.
The recent analysis of the properties of double magic nuclei~\cite{Brown:2013mga} 
puts more constraints,  $\varepsilon_S(\rho=0.1\mbox{\,fm}^{-3})=25.4\pm0.8$\,MeV.
An analysis of data on the neutron skin thickness of Sn isotopes and 
the isotope binding energy difference for heavy nuclei at slightly 
higher density value gave the result 
$\varepsilon_S(\rho=0.11\mbox{\,fm}^{-3})=26.65\pm0.2$\,MeV~\cite{Zhang:2013wna}.
For comparison, our results are presented in the last two columns of  
Table~\ref{table2}.
One can see that they are mainly consistent with the results in 
Refs.~\cite{Trippa:2008gr,Brown:2013mga,Zhang:2013wna}.

The discussions of density dependencies of the pressure 
in symmetric and asymmetric matter within 
the present approach can be found in  Ref.~\cite{Yakhshiev:2013eya}.
Here we only would like to note, that the modifications 
are also consistent with the results from other approaches.  
Moreover, the density dependencies of pressure in symmetric and 
asymmetric matter are weakly sensitive to the change of model parameters 
and the reproduced values correspond to the allowed region  
deduced from the experimental flow data and simulations studies by 
Danielewicz {\em et al.}~\cite{Danielewicz}.

In summary, the properties of nuclear matter at saturation density $\rho_0$ 
and  the extrapolations to the lower than $\rho_0$ regions are
found to be satisfactory and consistent 
with the results from other approaches. Therefore, we now concentrate 
on the extrapolations to the high density regions and discuss 
the properties of neutron stars.

\subsection{Properties of neutron stars}

Let us start from the mass-radius relations of neutron stars.
There are dramatic differences in predictions from the different 
approaches as concerned the mass-radius relation of neutron stars.
But most of the models with Equations of State of normal 
(nonstrange) nuclear matter predict existence of the regions 
around the one solar mass where the radius of neutron star is independent
from the mass.

\begin{figure*}[thb]
\vspace{0.2cm}
\hspace{.cm}
\centerline{\includegraphics[width=8cm]{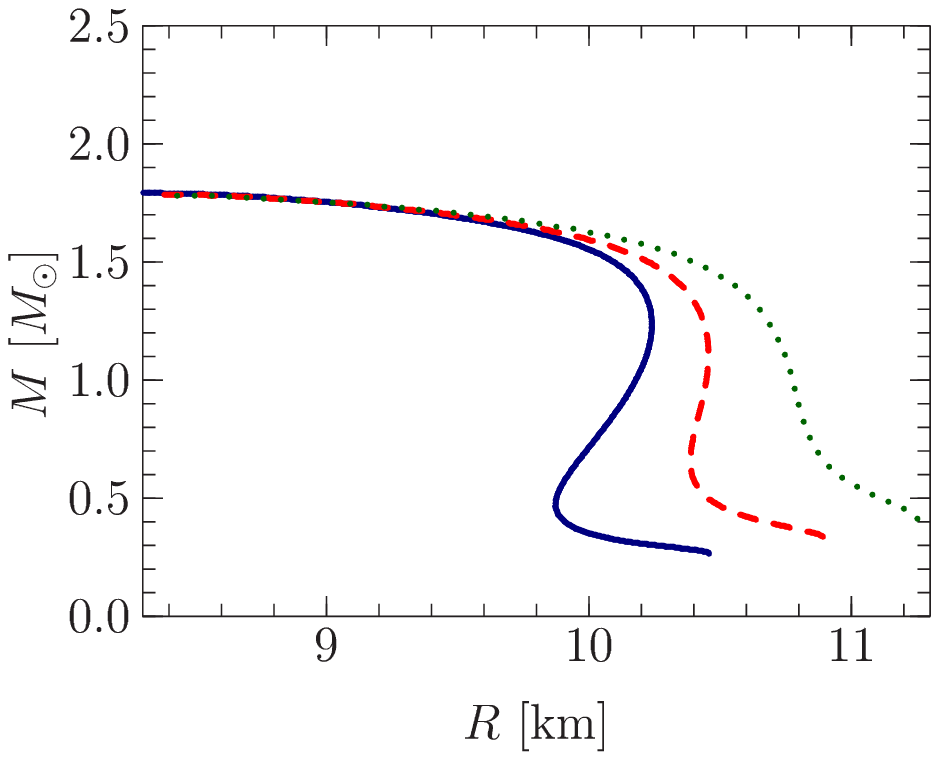}
\hspace{1.cm}\includegraphics[width=8cm]{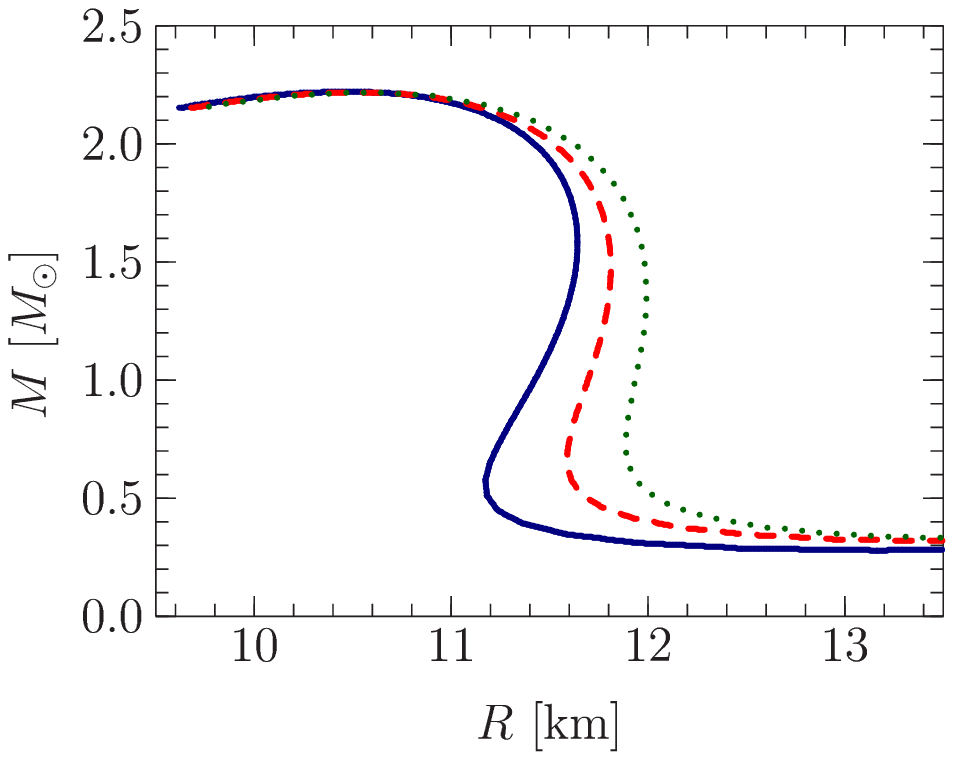}}
\caption{(Color online) The mass-radius relations of neutron stars
at the values of compressibility  
$K_0=250$~MeV (left panel, Set II) and $K_0=270$~MeV (right panel,
Set III). 
The value of symmetry energy at saturation density $\rho_0$ 
is chosen as $\varepsilon_S(1)=32$~MeV.
The solid, dashed and dotted curves correspond to the symmetry 
energy slop parameter $L$ values 30, 40 and 50~MeV. They represented 
by sets d,e and f in the Models II and III,
respectively (see Table~\ref{table2}).}
\label{fig:MR}
\end{figure*}

In the Fig.~\ref{fig:MR} we present our results 
corresponding to the two possible values of 
the compressibility of symmetric matter $K_0=250$\,MeV in the left 
panel and $K_0=270$\,MeV in the right panel, respectively.
One can see that the mass independent regions exist also in the 
framework of the present approach.
Another the common for many models
peculiarity, the existence of the maximal mass of the neutron star,
is also reproduced well in this model. As we mentioned above, the results 
are sensitive to the compressibility value,
if the compressibility value 
decreases it leads to decreased values of the maximal mass   
and radius of neutron star leading in such a way to more
compact neutron stars. From other side, 
the results are not sensitive to the change in the value of 
$\varepsilon_S(1)$ and, therefore, we presented only the results 
corresponding to the value of $\varepsilon_S(1)=32$\,MeV.
In general, our results are in qualitative agreement with the 
results from other approaches.

It is also interesting to compare our results in quantitative level
too. Measurements of the thermal spectra
from the quiescent low-mass X-ray binaries inside globular clusters
gave the possibility to fit the data sets with a neutron star
radius $R_{\rm NS}=9.1^{+1.3}_{-1.5}$~km at 90\% confidence 
level~\cite{Guillot:2013wu}. Determinations of the mass-radius relation, 
based on recent observations of both transiently 
accreting and bursting sources, gave the radius 
range between 10.4 and 12.9~km for 1.4 solar mass neutron 
stars~\cite{Steiner:2012xt}. 

In the present approach 
some class of parameters (Model III) defined in Table~\ref{table2}
give very good agreement with the above mentioned estimations in quantitative 
level. The properties of neutron stars reproduced using some 
subclasses of Model III is presented in Table~\ref{table3}.
\begin{table*}[thb]
\caption{Properties of the neutron stars from the 
different sets of parameters (see Tables~\ref{table1} 
and~\ref{table2} for the values of parameters):
$n_c$ is central number density, $\rho_c$ is central energy-mass density,
$R$ is radius of the neutron star, $M_{\rm max}$ is possible maximal mass,
$A$ is number of baryons in the star, $E_b$ is binding energy of the star.
In the left panel we represent the neutron star properties 
corresponding to the maximal mass $M_{\rm max}$ and in right panel 
approximately 1.4 solar mass
neutron star properties. The last two lines are 
results from the Ref.~\cite{Chabanat:1997qh}.
}
\begin{ruledtabular}
\begin{tabular}{c|cccccc|cccccc}
Set
&$n_c$
&$\rho_c$
& $R$
& $M_{\rm max}$
& $A$
& $E_{b}$
&$n_c$
&$\rho_c$
& $R$
& $M$
& $A$
& $E_{b}$\\
&[fm$^{-3}$]
&[$10^{15}$gr/cm$^{3}$]
&[km]
&[$M_\odot$]
&[$10^{57}$]
&[$10^{53}$erg]
&[fm$^{-3}$]
&[$10^{15}$gr/cm$^{3}$]
&[km]
&[$M_\odot$]
&[$10^{57}$]
&[$10^{53}$erg]\\
\colrule
III-a
&1.046&2.445&10.498&2.226&3.227&8.721
&0.479&0.861&11.587&1.402&1.898&3.503\\ 
III-b
&1.045&2.444&10.547&2.223&3.216&8.557
&0.471&0.861&11.772&1.402&1.895&3.453\\
III-c
&1.037&2.424&10.616&2.221&3.200&8.397
&0.460&0.832&11.953&1.402&1.887&3.339\\
III-d 
&  1.047 & 2.452  & 10.494 & 2.221 & 3.213 & 8.598 
& 0.481 & 0.867 & 11.619 & 1.402 & 1.893 & 3.422\\
III-e 
& 1.044 & 2.440 & 10.554 &2.218 & 3.203 & 8.495
& 0.473 & 0.858 & 11.809 & 1.403 & 1.890 & 3.384 \\
III-f 
& 1.040&2.433 & 10.609&2.216 & 3.189 &  8.311
& 0.464 & 0.842& 11.992&1.403 & 1.887 & 3.334\\
SLy230a~\cite{Chabanat:1997qh}
&1.15&2.69&10.25&2.10&2.99&7.07
&0.508&0.925&11.8&1.4&1.85&2.60\\
SLy230b~\cite{Chabanat:1997qh}
&1.21&2.85&\,\,9.99&2.05&2.91&6.79
&0.538&0.985&11.7&1.4&1.85&2.61\\
\end{tabular}
\end{ruledtabular}
\label{table3}
\end{table*}
One can see that, our results corresponding to more than 2$M_\odot$
as well as $\sim1.4M_\odot$ neutron stars are very similar to 
the estimations 
from the Refs.~\cite{Guillot:2013wu,Steiner:2012xt}.

For comparison, in Table~\ref{table3} we present two of the 
possible neutron stars parametrizations from the Ref.~\cite{Chabanat:1997qh}
as the representatives of the Skyrme effective forces
used in the density functional approach.
One can see that our results in qualitative agreement with the results 
from the Ref.~\cite{Chabanat:1997qh}.
It is also necessary to note, that the compressibility value used in 
Ref.~\cite{Chabanat:1997qh} is smaller $K_0=230$\,MeV in comparison 
with the compressibility value in Model III $K_0=270$\,MeV.
Therefore, we have more stiff EOS leading to slightly higher 
maximal mass and larger radius neutron stars. But all of the models 
presented in Table~\ref{table3} give more or less similar results 
as concerned 1.4$M_{\odot}$ mass neutron stars.

The central number densities of maximal mass neutron stars in the 
present approach are around $6.5\rho_0$
and the corresponding central mass-energy densities are around 
$2.44\times 10^{15}$\,gr/cm$^3$ ($\approx 1300$\,MeV\,fm$^{-3}$).
Our results are close to the results from nonrelativistic
potential model approaches discussed in Ref.~\cite{Lattimer:2000nx}.   

It is interesting also to compare the total baryon number of neutron star
in the present approach with the results from other approaches.
General Relativistic formula for the total baryon number of 
the neutron star is given by the integral
\begin{equation}
A=4\pi\int_0^R {\rm d}r\,r^2\rho(r)\left(1-\frac{2G{\cal M}(r)}{r}\right)^{-1/2}\,.
\end{equation}
One can find the radial dependence 
in the number density $\rho(r)$
from the relation $P=P(\rho/\rho_0)$ after finding 
the radial dependence of the pressure $P=P(r)$.
Our results are in qualitative agreement with results from 
Ref.~\cite{Chabanat:1997qh} (see Table~\ref{table3}).

After calculations of the total baryon number, one can also 
estimate the binding energy $E_b$ of the neutron star.
Due to the decrease of the gravitational mass $E_b$ is 
defined by the formula 
\begin{equation}
E_b=Am_N-M\,,
\end{equation}
where we used the mass of nucleon $m_N$ in free space.\footnote{Note,
that the authors of the Ref.~\cite{Chabanat:1997qh}
used 1/56 part of the ${}^{56}$Fe atom mass in calculating 
the binding energy of the neutron star.}
One can see that the calculated binding energies 
corresponding to 1.4 solar mass neutron stars  are consistent with 
the estimations made on the analysis of detected neutrinos from SN1987A: 
$E_b=3.8\pm 1.2\times 10^{53}$\,erg~\cite{SchaefferNature}.  

\begin{figure}[htb]
\vspace{0.2cm}
\hspace{.cm}
\centerline{\includegraphics[width=8cm]{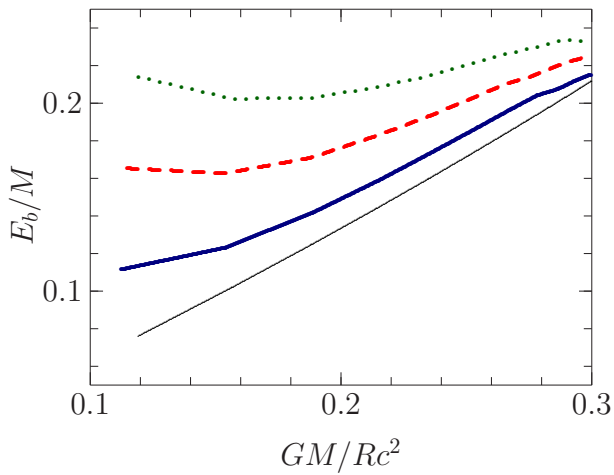}}
\caption{(Color online) Binding energy per unit mass 
$E_b/M$ of the neutron star
as a function of $GM/Rc^2$.
The solid, dashed and dotted curves correspond to the sets 
III-d, III-e and III-f defined in Table~\ref{table2}.
Thin and black solid curve represents an approximate relation Eq.~(\ref{LP}) 
presented in Ref.~\cite{Lattimer:2000nx}.}
\label{fig:BE}
\end{figure}

In addition we present the binding energy per unit mass versus 
$GM/Rc^2$ relation in Fig.~\ref{fig:BE} for some representatives 
of neutron star models given in Table~\ref{table3}.  
For comparison we represent also the outcome from
an approximate formula 
\begin{equation}
\frac{E_b}{M}\simeq \frac{0.6 GM}{Rc^2}\left(1-\frac{0.5GM}{Rc^2}\right)^{-1}
\label{LP}
\end{equation}
suggested by Lattimer and Prakash~\cite{Lattimer:2000nx}.
It is seen that in the present approach the small values of slop parameter 
$L_s\sim 30$\,MeV of symmetry energy give the results close to the parametrization 
in Eq.~(\ref{LP}).

\section{Summary}

In summary, we discussed the application of the in-medium 
modified chiral soliton model 
to the studies of asymmetric matter properties and neutron star structure.
The symmetric and asymmetric matter equations of state where reproduced by
very simple 5-parametric density approach to the single nucleon 
properties in the nuclear environment introducing the isospin 
breaking effects in the mesonic sector of the model. After reproducing the 
asymmetric matter properties near the saturation density 
of symmetric matter $\rho_0$  
we extrapolated the Equations of State to the high density 
and highly isospin asymmetric regions. Our primary goal was a
crude qualitative analysis of neutron star properties.
Nevertheless, at some given set of parameter values, 
our results are very close to the predictions from analysis 
of the data compiled during the observation of neutron 
stars~\cite{Guillot:2013wu,Steiner:2012xt}. 
In particular, 
the calculations showed that the properties of 
$\sim1.4M_\odot$ and $\sim2M_\odot$ neutron stars can be well reproduced within the present approach.

As an outlook for further studies we note that the in-medium chiral soliton 
model presented here can easily be extended to the studies of finite nuclei 
properties. This task can be realized in the local density approach
for the density of environment surrounding the soliton under the 
consideration. Those studies are under the way.

\begin{acknowledgments}
This work is supported by Basic Science Research Program through the National
Research Foundation (NRF) of Korea
funded by the Korean government (Ministry of Education, Science and
Technology -- MEST), Grant Number: 2011-0023478.
\end{acknowledgments}

\end{document}